\documentclass[twocolumn,preprintnumbers,amsmath,amssymb]{revtex4}

\usepackage{graphicx}
\usepackage{xspace}
\usepackage{epsf}
\usepackage{multirow}
\usepackage{amssymb}
\newcommand{\etal}{{\em et al}}

\begin{document}

\renewcommand{\etal}{{\em et al}}

\title{ab-initio study of different structures of CaC:
       Magnetism, Bonding, and Lattice Dynamics}

\author{Zahra Nourbakhsh}
\email{z.nourbakhsh@ph.iut.ac.ir}

\author{S. Javad Hashemifar}
\email{hashemifar@cc.iut.ac.ir}

\author{Hadi Akbarzadeh}

\affiliation{Department of Physics, Isfahan University of Technology,
         Isfahan, 84156-83111, Iran}

\begin{abstract}
  On the basis of ab-initio pseudopotential calculations,
  we study structural, magnetic, dynamical, and mechanical properties of the
  hypothetical CaC ionic compound in the rock-salt (RS), B2, zinc-blende (ZB),
  wurtzite (WZ), NiAs (NA), anti-NiAs (NA*), and CrB (B33) structures.
  It is argued that the ZB, WZ, NA, and RS structures are more ionic while the NA*, B2, and B33
  structures are more covalent systems.
  As a result of that, the nonmagnetic B33-CaC is the energetically preferred system,
  while the more ionic structures prefer a ferromagnetic ground state
  with high Fermi level spin polarization.
  The observed ferromagnetism in the more ionic systems is attributed to
  the sharp partially filled $p$ states of carbon atom in the system.
  In the framework of density functional perturbation theory,
  the phonon spectra of these systems are computed and the
  observed dynamical instabilities of the NA* and B2 structures are
  explained in terms of the covalent bonds between carbon atoms.
  The calculated Helmholtz and Enthalpy free energies indicate the highest stability
  of the B33 structure in a wide range of temperatures and pressures.
  Among the ferromagnetic structures, RS-CaC and ZB-CaC are reported, respectively, to be
  the most and the least metastable systems in various thermodynamics conditions.
  Several mechanical properties of the dynamically stable structures
  of CaC are determined from their phonon spectra.
\end{abstract}

\maketitle

\section{Introduction}

\begin{figure}[b]
  \includegraphics*[scale=0.9]{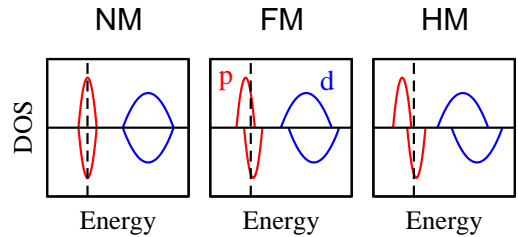}
  \caption{\label{stoner}
  Schematic spin resolved density of states (DOS) of  a $p$ magnetic
  system in the non-magnetic (NM), ferromagnetic (FM) and half-metallic (HM) states.
  The upward and downward plots show the spin majority and spin minority states
  and the dashed lines show the Fermi energy.}
\end{figure}

Using first-principle calculations, in 2004,
Kusakabe \etal.\cite{Kusakabe} predicted a new class of
$p$ ferromagnetic systems which are some hypothetical binary
ionic compounds with no transition metal, including CaP, CaAs and CaSb.
The $p$ magnetic materials exhibit significant exchange interaction in their valence $p$ electrons,
while in the conventional magnetic systems, $d$ or $f$ electrons
are responsible for the exchange interaction \cite{pmag}.
Further theoretical studies showed that this
new class of $p$ magnetic systems may exhibit half-metallicity
in the metastable zinc-blende (ZB) structure \cite{halfmetals}.
Half-metal ferromagnets are materials with perfect Fermi level spin polarization
and hence are promising as potential spin current sources in Spintronics devices.
Half-metallic ferromagnetism has been predicted in the ZB structure
of several II$^A$-IV$^A$, II$^A$-V$^A$, and I$^A$-V$^A$ compounds,
including CaC, CaN, and LiP \cite{Sieberer, cac, cac-2, cac-3}.
In all of these ionic binary compounds, the partially occupied
sharp $p$ band of the anion atom enhances the Stoner exchange
interaction and thus give rises to a ferromagnetic (FM) ground state.
This mechanism is schematically sketched in Fig.~\ref{stoner}.
The sharpness of the $p$ band is controlled by the cell volume,
system ionicity, and slight $p$-$d$ hybridization around the Fermi level \cite{Sieberer}.
As it is qualitatively argued in the figure, if exchange splitting is
larger than the band width, a half-metallic gap appears in the majority channel.
A major open issue in the novel $p$ magnetic compounds
is the stability of their hypothetical ferromagnetic structures.
First-principles computation of total energies and phonon spectra
may provide insightful information for understanding this open issue.

\begin{figure}
  \includegraphics*[scale=0.25]{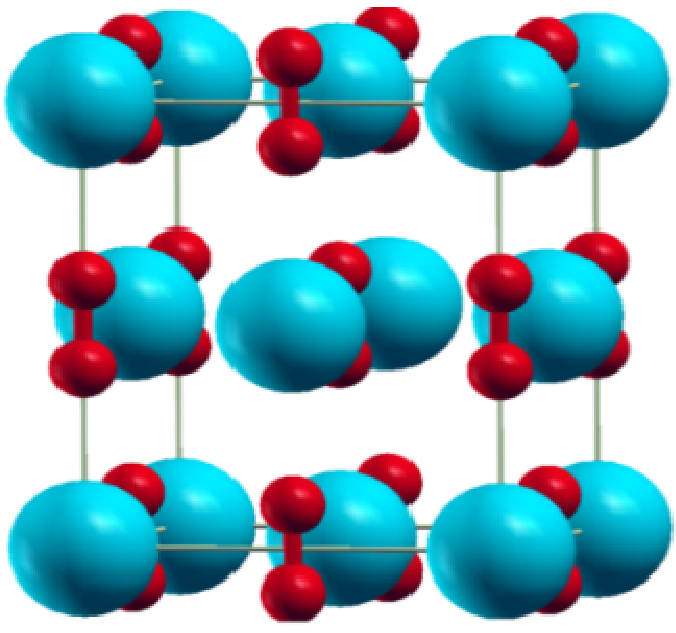}\hspace{2mm}
  \includegraphics*[scale=0.3]{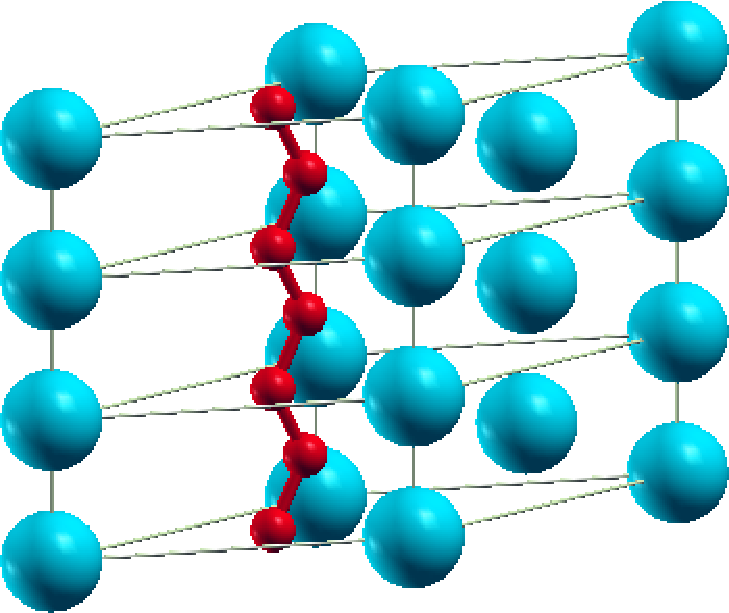}
  \caption{\label{b33-cac2}
 Left: the rock-salt structure of CaC$_2$, composed of Ca ions (big blue balls) and 
 carbon dimers (small red balls). 
 Right: the monoclinic B33 lattice of CaC ($a=b$ and $\alpha=28^{\circ}$), 
 composed of two CaC planes ($z=0$ and $z=1/2$) per cell (three unit cells are visualized). 
}
\end{figure}

In this paper, we focus on the hypothetical CaC compound as a potential
$p$ magnetic system and apply density functional calculations
to investigate its structural, magnetic, and dynamical properties in seven different structures:
monoclinic CrB (B33),
cubic rock-salt (RS), CsCl (B2) and ZB,
and hexagonal wurtzite (WZ), NiAs (NA) and
anti-NiAs (NA*) structures.
The NA* structure is made by permuting the Ni and As atoms in the NA structure.
Since many alkaline earth silicide and germanide compounds,
like CaSi and CaGe, crystalize in the CrB type lattice \cite{b33, b33-2, b33-3},
this structure was included in our search for the stable structure of CaC (See Fig.~\ref{b33-cac2}.).
The well-known compound of Ca and C is CaC$_2$ that is a nonmagnetic insulator
and crystallizes in the rock-salt structure in which C$_2^{2-}$ dimers
and Ca$^{2+}$ ions occupy the 4a and 4b Wyckoff positions (See Fig.~\ref{b33-cac2}.) \cite{cac2}.
Understanding the $p$ ferromagnetism in the hypothetical CaC compound
may provide new opportunities for developing ferromagnetic carbon based nano-structures.

After reviewing our computational methods in the next section,
the structural and magnetic properties of all selected CaC structures
will be discussed in section III.
Then the phonon spectra of the systems will be presented to
investigate the dynamical stability and elastic properties of the CaC structures.
Our conclusions are presented in the last section.

\section{Computational Methods}

The geometry optimizations and electronic structure calculations were
carried out by using the PWscf code of the {\sc Quantum-Espresso} package \cite{pwscf}.
The generalized gradient approximation (GGA) in the scheme of Perdew,
Burke, and Ernzerhof (PBE)\cite{pbe} and norm-conserving pseudopotentials \cite{pseudo},
by considering 10 valence electrons for Ca and 4 valence electrons for C,
were used throughout our calculations.
The Kohn-Sham single particle wave functions were expanded in plane waves up to energy
cutoff of 80 Ry and the Fourier expansion of electron density were cut at 320 Ry.

The Brillouin zone integrations were performed on the optimized $\Gamma$-centered
symmetry-reduced Monkhorst-Pack k meshes using the Methfessel-Paxton\cite{mp} smearing
method with a broadening parameter of 1 mRy.
All computational parameters were optimized to achieve
total energy accuracy of about 0.1 mRy/formula unit ($fu$).
The electronic structure calculations were performed in the scalar
relativistic limit, ignoring the relativistic spin-orbit interaction
which is expected to be small in the light carbon and calcium atoms.
The internal parameters in the WZ and B33 structures were accurately relaxed.
The phonon frequencies are calculated using the density functional
linear response method for metallic systems \cite{dfpt}.

For accurate description of bonding and magnetism in the system,
we employ topological analysis of electron charge density \cite{tecd},
based on Bader's theory of atoms in molecules \cite{Bader}.
In this scheme, bond points are defined as the saddle points
of electron density between neighboring nuclei.
At the bond saddle points, electron density displays a minimum
in the bond direction and two maxima in the perpendicular directions.

\section{Structural and Magnetic Properties}

In order to find the structural properties of CaC in the selected structures,
we computed the primitive cell energy of all structures
as a function of their primitive volume,
in the ferromagnetic and nonmagnetic phases.
The results indicate that the B2, NA* and B33 structures of CaC are
nonmagnetic around their equilibrium volume,
while other systems (ZB, WZ, NA, and RS)
stabilize in a ferromagnetic ground state.
Moreover, the nonmagnetic structures were found to be more stable
than the ferromagnetic systems, and among them B33 exhibits the lowest energy.

\begin{figure}
  \includegraphics*[scale=0.8]{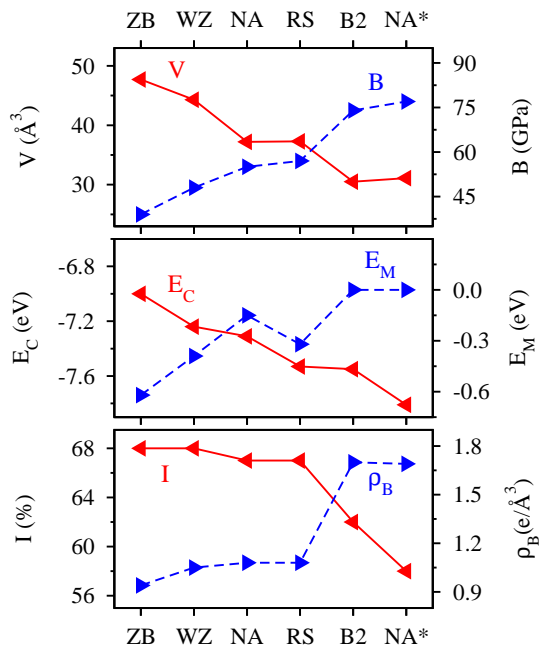}
  \caption{\label{struct}
  Calculated equilibrium properties of CaC in the selected structures.
  $V$: equilibrium primitive volumes, $B$: bulk modulus, $E_C$: cohesive energy per formula unit,
  $E_M$: magnetization energy defined as the difference between
  the ferromagnetic and nonmagnetic cohesive energies,
  $I$: ionicity, $\rho_B$: total electron density at bond points.}
\end{figure}

The calculated equilibrium properties, determined by
fitting the Murnaghan equation of state \cite{Murnaghan}
to the calculated energy-volume data, are presented in Fig.~\ref{struct}.
It is seen that in the sequence of ZB, WZ, NA, RS, B2, NA*, and B33 structures,
the cohesive energy, equilibrium volume, and compressibility
(inverse of the bulk modulus) are decreasing.
These behaviors indicate enhancement of the bond strength and bond stiffness in this sequence.
For better understanding, the ionicity parameter $I$ and
the total electron density at bond points $\rho_B$ (per formula unit)
were also computed and plotted in Fig.~\ref{struct}.
The ionicity parameter, being defined as the transferred charge
from cation to anion normalized by the cation valence charge,
is a measure of the ionic bonding in the system.
On the other hand, $\rho_B$ is the sum of electron densities
at all bond points (per formula unit)
and thus speculates the strength of the covalent bonding in the system.
We observe that the enhancement of the bond strength and
stiffness in the studied structures is accompanied by
decreasing the ionic and increasing the covalent bonding in the systems.
It is consistent with the general believe that the directional covalent bonds
form solids with high strength and high stiffness.
The calculated results (Fig.~\ref{struct}) indicate that the nonmagnetic
B33 structure exhibits the highest covalency and the lowest ionicity
among the studied systems.

\begin{table*}
  \newcommand{\rb}{$\rho_b$}
  \newcommand{\dhf}{$\Delta H_f$}
  \newcommand{\nb}{$N_b$}
  \newcommand{\moment}{\multicolumn{3}{c}{magnetic moments}}
  \newcommand{\CC}{\multicolumn{3}{c}{C~$-$~C}}
  \newcommand{\CCa}{\multicolumn{3}{c}{C~$-$~Ca}}
  \caption{\label{eq-tab}
  Computed formation energies \dhf~(eV/$fu$),
  equilibrium lattice parameters $a$ and $c$~(\AA),
  equilibrium lattice volume per formula unit $V$~(\AA),
  atomic and total magnetic moments ($\mu_B$),
  and C-C and C-Ca bond properties in the selected structures of CaC;
  $P$(\%): total spin polarization at the Fermi level,
  $d$(\AA): equilibrium distance,
  $N_b$: number of the bonds per formula unit, and
  $\rho_b$ ($e/$\AA$^3$): bond electron density.
  The WZ structure involves slightly different vertical and in-plane
  bonds and hence, in this case, the average bond properties are reported.
  The equilibrium value of $cos(\alpha)$ in the B33 structure were found to be about 0.886.}
  \begin{ruledtabular}
  \begin{tabular}{lcccccccccccccccc}
       &      &     &  &      &     &      \moment       &&        \CC      &&      \CCa       \\
                                       \cline{7-9}          \cline{11-13}     \cline{15-17}
       & \dhf & $a$  & $c$  &$V$& $P$ &   C  &  Ca  & total&&  $d$ &\nb&  \rb &&  $d$ &\nb& \rb  \\
    \hline
    B33&  0.5 & 7.13 & 2.76 &31.7& 0   &  0   &   0  & 0.00 && 1.55 & 1 & 1.42 && 2.68 & 6 & 0.14 \\
    NA*&  1.8 & 4.19 & 4.06 &31.0& 0   &  0   &   0  & 0.00 && 2.03 & 1 & 0.49 && 2.62 & 6 & 0.20 \\
    B2 &  2.1 & 3.13 & ---  &30.7& 0   &  0   &   0  & 0.00 && 3.13 & 3 & 0.14 && 2.71 & 8 & 0.16 \\
    RS &  2.1 & 5.29 & ---  &37.1& 70  & 1.77 & 0.20 & 1.97 && 3.74 & 0 &  --- && 2.65 & 6 & 0.18 \\
    NA &  2.3 & 3.64 & 6.48 &37.2& 70  & 1.79 & 0.19 & 1.98 && 3.64 & 0 &  --- && 2.65 & 6 & 0.18 \\
    WZ &  2.4 & 4.38 & 5.30 &44.1& 90  & 1.80 & 0.19 & 1.99 && 3.66 & 0 &  --- && 2.58 & 5 & 0.20 \\
    ZB &  2.6 & 5.75 & ---  &47.5& 100 & 1.82 & 0.18 & 2.00 && 4.07 & 0 &  --- && 2.49 & 4 & 0.24 \\
  \end{tabular}
  \end{ruledtabular}
\end{table*}

We observe that the ferromagnetic structures have higher ionicity
and lower covalency, compared with the nonmagnetic structures.
These observations evidence that the exchange interaction in the CaC structures
is significantly enhanced by the ionicity of the system.
The reason is that the ionic bonding between Ca and C atoms transfers
(nominally) the 2 electrons of the Ca 2$s$ orbital to the C 2$p$ orbital,
leaving a partially occupied 2$p^4$ orbital at the Fermi level of the system.
In the absence of a strong covalent bonding between atoms, this partially
occupied orbital remains sharp at the Fermi level and hence enhances
the stoner exchange interaction in the system.

The formation energy, equilibrium lattice parameters, Fermi level spin polarization, bond lengths,
and atomic magnetic moments of the seven studied structures of CaC
are listed in table~\ref{eq-tab}.
The presented formation energies ($\Delta H_f$)
are defined as the difference between the cohesive anergy of bulk CaC
and sum of the cohesive energies of graphite and bulk fcc calcium \cite{coh}.
The obtained formation energies are all positive, indicating endothermic formation
of bulk CaC from graphite and bulk calcium.
After topological analysis of charge densities,
we identified that, in addition to the carbon-calcium
bonding in all systems, there are some covalent carbon-carbon bonds in the nonmagnetic structures.
The number of these C-C bonds in the systems, their length, and their electron
density are also listed in table~\ref{eq-tab}.
In order to have a rough estimation of the strength of the covalent C-C bonds
in the systems, we calculated the equilibrium properties of diamond
and found a C-C bond length of 1.55 \AA~
and bond density of 1.56 $e$/\AA$^3$ in this system.
Hence, it is seen that in the B33 structure, the C-C bonds seem to be as strong as
the C-C bonds in diamond. This observation explains the more stability of the B33
structure of CaC, compared with the other studied structures.
In the nonmagnetic NA* structure, covalent bonding between C atoms in the $z$ direction
significantly decreases the c/a ratio, compared with the NA structure.

For better understanding of the results, we note that while
the C-Ca bonding is mainly ionic, the C-C bonding is totally covalent.
Observation of the covalent C-C bonds in the nonmagnetic B33,
NA*, and B2 structures explains the high covalency of these systems (Fig.~\ref{struct}).
Moreover, the slight increasing of the magnetic moment in RS, NA, WZ, and ZB series
is generally accompanied by increasing (decreasing) the C-C (C-Ca) distance.
Decreasing the C-Ca distance intensifies the ionic bonding in the system
and hence enhances the magnetic moment via increasing the charge transferred
from the Ca 2$s$ orbital to the partially filled C 2$p$ orbital (Fig.~\ref{struct}).
In contrary, the reduction of the C-C distance enhances
the covalent bonding in the system and consequently increases
the valence band width and then decreases the magnetic moment.
These findings, further evidence the strong impact of ionic bonding
 and lattice parameter on the ferromagnetic behavior of CaC.

All magnetic structures show a magnetic moment of about
2~$\mu_B/fu$ with no covalent C-C bond.
In the ZB structure, the magnetic moment is exactly equal to 2~$\mu_B/fu$
which is due to the half-metallic behavior of this system argued by the calculated
perfect Fermi level spin polarization (table \ref{eq-tab}).
We observe that the magnetic moments are mainly carried by the C atoms
and the Ca atoms have a small moment parallel to the C moment.
The dominant magnetic contribution of carbon comes from the strong ionic C-Ca
bonding which highly evacuates the 2 valence electrons of the Ca 2$s$ shell.
The empty spin down C 2$p$ orbital gives rises to more charge transferring
in the minority channel and hence leaving a positive moment on the Ca atom.
All FM structures of CaC have a perfect or nearly semiconducting majority channel
and the Fermi level spin polarizations come from the minority electrons.
Hence it seems that these systems are capable of conducting
spin currents antiparallel to their magnetic moment.

\begin{figure}
  \includegraphics*[scale=0.8]{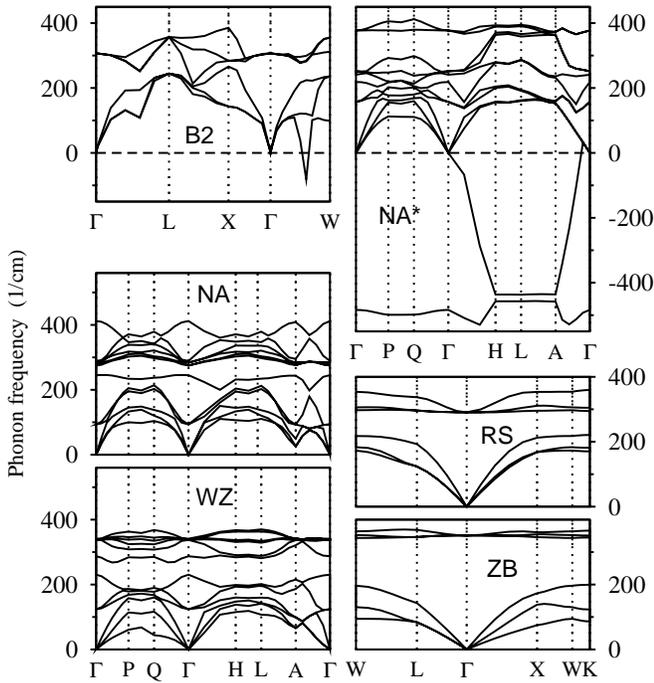}
  \caption{\label{ph-band}
  Calculated phonon band structure of CaC in the B33, B2, RS, ZB, NA*,
  NA, and WZ structures. The bands are plotted in comparable scales.}
\end{figure}

\section{Dynamical Properties}

The dynamical properties of CaC in the cubic RS, ZB, B2,
hexagonal WZ, NA, NA*, and monoclinic B33 structures are studied by using
their calculated phonon spectra in the framework of density functional perturbations theory \cite{dfpt}.
The obtained phonon band structures in the equilibrium
ground states are displayed in Fig.~\ref{ph-band}.
The negative values in these plots correspond to the imaginary phonon frequencies.
It is observed that the B2 and NA* structures have some imaginary phonon frequencies
and hence are dynamically unstable.
The dynamical instability of the B2 structure occurs in some
transverse acoustic phonon modes in a small portion of the BZ
in the [110] direction while NA*-CaC has a broad instability coming from
an optical phonon band in the whole BZ and a longitudinal acoustic band
in the [0001] direction.

\begin{figure}
  \includegraphics*[scale=0.7]{ph-pdos}\hspace{2mm}
  \includegraphics*[scale=0.3]{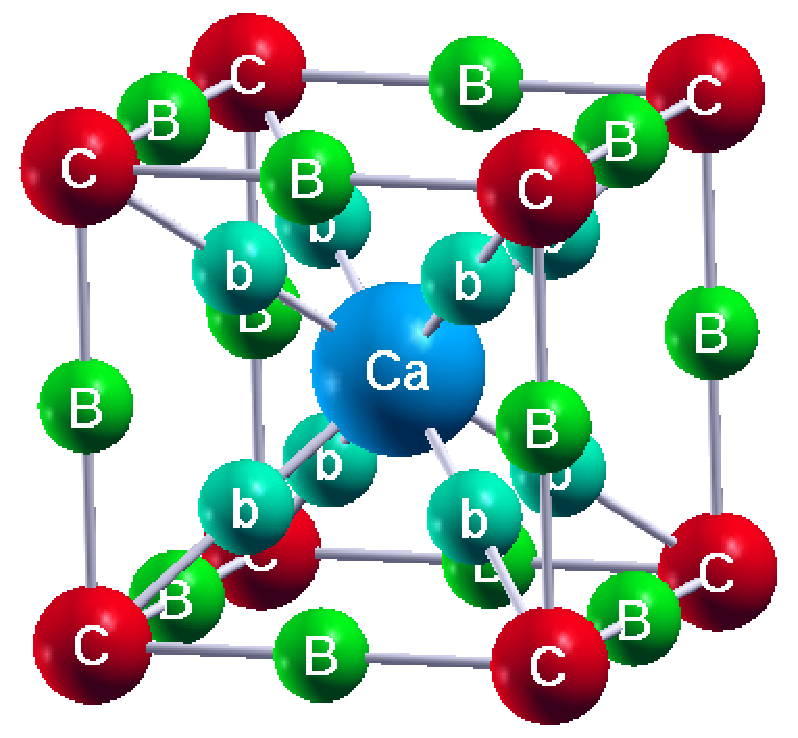}
  \caption{\label{ph-pdos}
  Left: Total phonon DOS (solid line) and carbon contribution to
  the phonon DOS (shaded area) in the B2 and NA* structures of CaC.
  The phonon DOS of the B2 structure in the negative (imaginary)
  frequencies is enlarged 50 times (inset) to be more visible.
  Right: Calculated topological bonds in B2-CaC,
  The green balls with letters b and B stand for the ionic Ca-C and
  covalent C-C bond points, respectively.}
\end{figure}

In order to understand the origin of the dynamical instabilities,
we calculated the atom resolved phonon density of states of the B2 and NA*
structures (Fig.~\ref{ph-pdos}).
It is clearly seen that the imaginary phonon frequencies
mainly come form carbon oscillations.
This observation indicates that the dynamical instability of the B2 and NA*
structures is due to the tendency of their carbon atoms for reconfiguration.
Existence of the carbon dimers in the CaC$_2$ compound \cite{cac2},
provides an experimental evidence for the tendency of carbons in CaC
to get close together and dimerise.
This statement, is consistent with the observed covalent bonds between carbon atoms
in the B2 and NA* structures (table \ref{eq-tab}).
The covalent bonds of the B2 structure, as a prototype, are visualized in Fig.~\ref{ph-pdos}.
As it is visible in the figure, the coplanar covalent carbon-carbon bonds in the B2 structure
may evidence the tendency the carbon atoms to reconfigure in the diagonal [110]
(and other equivalent) directions.
It is consistent with the observed dynamical instability of B2-CaC
in the [110] $\Gamma$W direction (Fig.~\ref{ph-band}).
Similar arguments apply to the phonon instabilities of NA*-CaC.

\begin{figure}
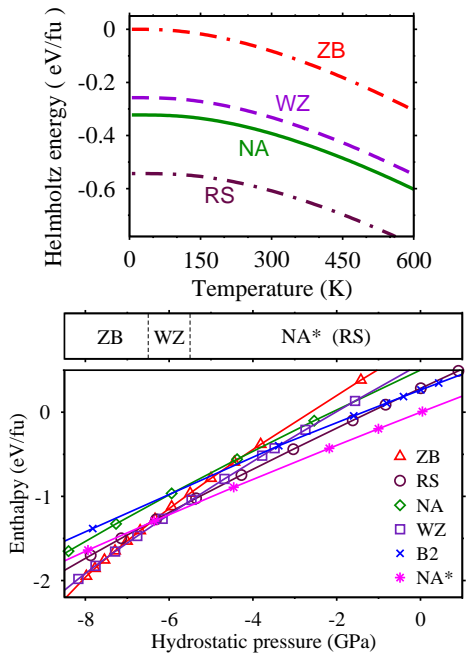

  \includegraphics*[scale=0.7]{helmholtz}
  \includegraphics*[scale=0.7]{enthalpy}
  \caption{\label{free-energy}
  Calculated Helmholtz and enthalpy free energy of the dynamically
  stable structures of CaC as a function of temperature and hydrostatic pressure.}
\end{figure}

\begin{table*}
  \caption{\label{elastic}
  Calculated mechanical properties of CaC in the B33, RS, NA, WZ, and ZB structures,
  $T_D$(K): Debye temperature, $C_{ij}$(GPa): elastic constants, $B_{ph}$, $Y_{ph}$, and $G_{ph}$(GPa):
  bulk, Young, and shear moduli calculated from the elastic constants,
  $B$ and $Y$(GPa): directly calculated bulk and Young moduli from the
  energy-volume and energy-lattice parameter curves, and $\sigma$: poison ratio.}
  \begin{ruledtabular}
  \begin{tabular}{lcccccccccccccc}
       &$T_D$&$C_{11}$& $C_{33}$ & $C_{44}$ & $C_{66}$ & $C_{12}$ & $C_{13}$&
    $B_{ph}$& $B$ & $Y_{ph}$ & $Y$ & $G_{ph}$ & $B/G$ & $\sigma$   \\
    \hline
    B33& 900 &     &     & 40 & 20 &     &    &    & 84 &     & 241 & 118 & 0.71 & 0.02 \\
    RS & 470 & 153 & 153 & 58 & 58 &  13 & 13 & 60 & 60 & 139 & 124 &  63 & 0.95 & 0.08 \\
    NA & 466 & 131 & 196 & 55 & 72 & -14 & 13 & 54 & 57 & 138 & 149 &  64 & 0.84 & 0.05 \\
    WZ & 459 & 76  & 200 & 10 & 13 &  49 & -2 & 49 & 48 &  68 &  61 &  27 & 1.81 & 0.27 \\
    ZB & 458 & 53  &  53 & 20 & 20 &  36 & 36 & 42 & 42 &  42 &  44 &  16 & 2.63 & 0.40 \\	
  \end{tabular}
  \end{ruledtabular}
\end{table*}

The phonon spectra of materials is a rich source of
mechanical and thermodynamical information.
The vibrational based thermodynamic properties of a system is studied
by calculating the vibrational contribution to the
Helmholtz free energy ($F$) as follows:
$$
F(T)=E-k_BT\int d\omega D(\omega)
     \ln(\frac{e^{-\beta\hbar\omega/2}}{1-e^{-\beta\hbar\omega}})
$$
where $E$ is the total energy of the system, the integration is over
the phonon frequencies, $D(\omega)$ is the phonon DOS,
$k_B$ is the Boltzmann constant, $T$ is the kelvin temperature, and $\beta=1/k_BT$.
By inserting the calculated phonon DOS into the above equation,
the Helmholtz free energy of CaC in its five stable structures
were calculated and presented in Fig.~\ref{free-energy}.
The results indicate stability of the B33 structure in the whole temperature range.
The relative stability of the studied systems at
elevated pressures, should be investigated in terms of the enthalpy free energy: $H=E-pV$,
where $E$ is the converged primitive cell energy, $V$ is the primitive volume,
and $p$ is the hydrostatic pressure calculated from the slope of the energy-volume curve.
The obtained enthalpy free energy of the dynamically stable structures, presented in Fig.~\ref{free-energy},
indicates thermodynamic stability of the B33 structure in a wide range of hydrostatic pressures.
Among the studied ferromagnetic structures, it is seen that RS-CaC exhibits lower free energy
in a broad range of temperatures and pressures.

Finally, we have used our first-principle phonon spectra to calculate
various mechanical properties of the stable structures of CaC (table~\ref{elastic}).
The Debye temperature ($T_D$), which measures hardness of a solid,
was calculated by using the average values of speed of sound in different directions
and density of atoms in the system \cite{deby}.
The elastic constants are calculated from the slopes of the acoustic
phonon bands at the $\Gamma$ point \cite{elastic}.
For the monoclinic B33 structure, finding all elastic constants are a complicated task,
as this structure has more independent elastic constants which should be calculated
from the behaviour of the system in the less symmetric directions;
hence in this case only two diagonal elastic constants were calculated.
The bulk ($B$) and shear ($G$) moduli in the cubic and hexagonal structures
are directly calculated from the elastic constants \cite{elastic1},
and then the Young modulus ($Y$) is calculated by $Y=9BG/(3B+G)$.
As it was mentioned, in the B33 structure only two
elastic constant were calculated which are not enough for calculation of elastic moduli,
hence the bulk and Young moduli of this system were only determined from direct calculation.
It is observed that the phonon based
bulk and Young moduli are close to the directly calculated values (table~\ref{elastic}).
This consistency confirms reliability and accuracy of our phonon spectra calculations.
In the hexagonal structures, two schemes proposed by Reuss and Voigt \cite{elastic1},
were applied for calculating the bulk and shear moduli from the elastic constants.
In the NA structure, we found that the Reuss scheme gives
better bulk and Young moduli, compared with the directly calculated values,
while in the hexagonal WZ structure, the Voigt scheme exhibits better performance.

The $B/G$ ratio, presented in table~\ref{elastic}, qualitatively
determine the ductility/brittleness ratio of a material \cite{BG}.
Therefore, among these CaC structures, the most ductile and brittle systems are expected to be
the ZB and B33 structures, respectively.
It is consistent with the observed strong covalent bonds in B33, evidencing the hardness of this structure,
similar to the diamond that is a very hard and thus brittle material.
The mechanical parameter poison ratio, calculated by $\sigma=1/2(1-Y/3B)$,
may be qualitatively used to compare structural stiffness of materials;
more stiff materials have lower poison ratio.
Hence the B33, RS and NA structures of CaC are expected to
be more stiff systems, compared with the WZ and ZB structures,
in general agreement with the observed bond electron densities (Fig. \ref{struct}).

\section{Conclusions}

In this paper we employed density functional-pseudopotential calculations
to investigate structural, magnetic, dynamical, and mechanical properties of binary compound CaC
in the zinc-blende (ZB), wurtzite (WZ), rock-salt (RS), B2, NiAs (NA), anti-NiAs (NA*),
and B33 structures.
Accurate investigation of the atomic charges and bond points of the structures
showed that the covalency of the system increases along
the ZB-WZ-NA-RS-B2-NA*-B33 sequence, while the system ionicity decreases.
As a result of that, the B33 structure exhibit the highest stability along with
a nonmagnetic ground state while the higher energy ZB, RS, WZ, and NA structures
stabilise in ferromagnetic states.
The observed $p$ ferromagnetism in the ferromagnetic structures of CaC
was argued to be due to the strong ionic bonding in the system
which significantly enhances the sharpness of the partially
filled $p$ band at the Fermi level in support of the stoner exchange interaction.
Calculated phonon spectra indicates dynamical instability of the nonmagnetic NA* and B2 structures
which was attributed to the observed tendency of the carbon atoms of the systems
toward reconfiguration.
The nonmagnetic B33 structure of CaC, among the studied structures,
was argued to be the most stable, most stiff, and most brittle system which
retains its stability over a broad range of temperatures and hydrostatic pressures.
Among the ferromagnetic structures, RS-CaC exhibits the highest metastability
over a broad range of of pressures and temperatures,
while the half-metallic ZB structure is found to be the least metastable system.

\section*{Acknowledgment}

This work was jointly supported by the Vice Chancellor of Isfahan
University of Technology (IUT) in research affairs, ICTP affiliated
centre in IUT, and Centre of Excellence for Applied Nanotechnology.
SJH acknowledges the Abdus Salam International Center for Theoretical
Physics (ICTP) for supporting his visit to ICTP.


\begin{thebibliography}{24}
\bibitem[{coh()}]{coh}
\bibinfo{note}{The cohesive energy of graphite and fcc calcium was calculated
  with the same accuracy as CaC calculations. In the case of fcc Ca, the volume
  optimization was performed while in graphite the $c$ lattice parametetr were
  set to the experimentalr value and the other lattice parameter was optimized.
  The optimized lattice parameters agree with the exprimental values with the
  error of less than $0.01$ \AA.}
\end{thebibliography}
\end{document}